\documentclass{ws-procs9x6}
\usepackage{wrapfig}
\usepackage{epsfig}

\let\ensm=\ensuremath
\newcommand{\jipsi}{\ensm{{\rm J}\!/\Psi}}

\def\Pp{{\rm p}}

\newcommand{\Pb}[1]{\ensm{\mkern 1mu\overline{\mkern -1mu{\rm #1}\mkern -2mu}\mkern 2mu}}

\def\Pbp{\Pb{p}}

\newcommand{\be}{\begin{equation}} \newcommand{\ee}{\end{equation}}
\newcommand{\ba}{\begin{eqnarray}} \newcommand{\ea}{\end{eqnarray}}

\begin{document}
\title{
Transversity in Drell-Yan process of polarized protons and
antiprotons in PAX experiment\footnote{ \uppercase{T}his work is
partially supported by grants \uppercase{INTAS} 00/587,
\uppercase{RFBR} 03-02-16816 and \uppercase{DFG-RFBR} 03-02-04022
and by \uppercase{V}erbundforschung of \uppercase{BMBF}.}}
\author{\underline{A.~V.~EFREMOV}$^a$, K.~GOEKE$^b$ and P.~SCHWEITZER$^b$}
\address{$^A$Joint Institute for Nuclear Research,
        Dubna, 141980 Russia\\
$^B$Institut f\"ur Theoretische Physik II,
        Ruhr-Universit\"at Bochum,  D-44780 Bochum, Germany}
\maketitle

\abstracts{
Estimates are given for the double spin asymmetry in
lepton-pair production from collisions of transversely polarized
protons and antiprotons for the kinematics of the recently
proposed PAX experiment at GSI on the basis of predictions for
the transversity distribution from the chiral quark soliton
model.
}

\paragraph*{1. Introduction.}
The leading structures of the nucleon in deeply inelastic
scattering processes are described in terms of three twist-2
parton distribution functions (PDF) -- the unpolarized
$f_1^a(x)$, helicity $g_1^a(x)$, and transversity $h_1^a(x)$.
Owing to its chirally odd nature $h_1^a(x)$ escapes measurement
in DIS experiments. The transversity was originally introduced in
the description of the Drell-Yan process of transversely
polarized protons\cite{Ralston:ys}. Alternative processes have
been discussed also, e.g. the Collins effect in semi-inclusive
deeply inelastic scattering experiments at HERMES, CLAS and
COMPASS could be (partly) understood in terms of
transversity\cite{Efremov:2001cz}. However, in all these
processes $h_1^a(x)$ enters in connection with some badly
known\cite{todd} Collins fragmentation function. Moreover these
processes involve the introduction of transverse parton momenta,
and for none of them a strict factorization theorem could be
formulated so far. So, the Drell-Yan process remains up to now
the theoretically cleanest and safest way to access $h_1^a(x)$.

The first attempt to study $h_1^a(x)$ by means of the Drell-Yan
process is planned at RHIC. Dedicated estimates, however,
indicate that at RHIC the access of $h_1^a(x)$ by means of the
Drell-Yan process is very difficult. The main reason is that the
observable double spin asymmetry $A_{TT}$ is proportional to a
product of transversity quark and antiquark PDF. The latter are
small, even if they were as large as to saturate the Soffer upper
limit.

This problem can be circumvented by using an antiproton beam.
Then $A_{TT}$ is proportional to a product of transversity quark
PDF from the proton and transversity antiquark PDF from the
antiproton (which are equal due to charge conjugation). Thus in
this case one can expect sizeable counting rates. The challenging
program how to polarize an antiproton beam has been recently
suggested in the PAX experiment at GSI\cite{PAX}. The technically
realizable polarization of the antiproton beam more
than\cite{Rathmann:2004pm} $(5-10)\%$ and the large counting
rates make the program rather promising.

In the talk I shortly describe our quantitative estimates for the
Drell-Yan double spin asymmetry $A_{TT}$ in the kinematics of the
PAX experiment at LO QCD. (For more details and references
see\cite{Efremov:2004qs}.) We also will estimate the recently
suggested analog double spin asymmetry in $\jipsi$
production\cite{Anselmino:2004ki}. For the transversity
distribution we shall use predictions from the chiral quark
soliton model\cite{Schweitzer:2001sr}. This model was derived
from the instanton model of the QCD vacuum\cite{Diakonov:2002fq}
and describes numerous nucleonic properties without adjustable
parameters to within $(10-30)\%$ accuracy. The field theoretic
nature of the model allows to consistently compute quark and
antiquark PDF which agree with
parameterizations\cite{Gluck:1994uf} to within the same accuracy.
This gives us a certain confidence that the model describes
$h_1^a(x)$ with a similar accuracy.

\paragraph*{2. Lepton pair production in {\boldmath $\Pp\uparrow\Pbp\downarrow$}.}
The process $\Pp\Pbp\to \mu^+\mu^-X$ can be characterized by the
invariants: Mandelstam $s=(p_1+p_2)^2$ and dilepton invariant
mass $Q^2=(k_1+k_2)^2$, where $p_{1/2}$ and $k_{1/2}$ are the
momenta of respectively the incoming proton-antiproton pair and
the outgoing lepton pair, and the rapidity $y=\frac12\,{\rm
ln}\frac{p_1(k_1+k_2)}{p_2(k_1+k_2)}$. The double spin asymmetry
in Drell-Yan process is given by
\be\label{Eq:ATT-0}
    \frac{N^{\uparrow\uparrow}-N^{\uparrow\downarrow}}
    {N^{\uparrow\uparrow}+N^{\uparrow\downarrow}}
    = D_P \; \frac{\sin^2\theta}{1+\cos^2\theta}\;\cos 2\phi\;
    A_{TT}(y,Q^2) \;,
\ee
where $\theta$ is the emission angle of one lepton in the
dilepton rest frame and $\phi$ its azimuth angle around the
collision axis counted from the polarization plane of the hadron
whose spin is not flipped in Eq.~(\ref{Eq:ATT-0}). The factor
$D_P$ takes into account polarization effects. At LO QCD $A_{TT}$
is given by \be \label{Eq:ATT-1} A_{TT}(y,Q^2)=\frac{\sum_a e_a^2
h_1^a(x_1,Q^2) h_1^a(x_2,Q^2)} {\sum_b e_b^2 f_1^b(x_1,Q^2)
f_1^b(x_2,Q^2)}\;, \ee where the parton momenta $x_{1/2}$ in
Eq.~(\ref{Eq:ATT-1}) are $x_{1/2} = \sqrt{\frac{Q^2}{s}}\,e^{\pm
y}$. In Eq.~(\ref{Eq:ATT-1}) use was made of the charge
conjugation invariance.

In the PAX experiment an antiproton beam with energies in the
range $(15-25)\,{\rm GeV}$ could be available, which yields
$s=(30-50)\,{\rm GeV}^2$ for a fixed proton target. The region
$1.5\,{\rm GeV} < Q < 3\,{\rm GeV}$, i.e. below the $J/\Psi$
threshold but well above $\Phi(1020)$-decays (and with
sufficiently large $Q^2$) would allow to explore the region
$x>0.2$. However, in principle one can also address the resonance
region itself and benefit from large counting
rates\cite{Anselmino:2004ki} since the unknown $q\bar q J/\Psi$
and $J/\Psi\mu^+\mu^-$-couplings cancel in the ratio in
Eq.~(\ref{Eq:ATT-0}) as argued in Ref.\cite{Anselmino:2004ki}.
Keeping this in mind we shall present below estimates for
$s=45\,{\rm GeV}^2$, and $Q^2=5\,{\rm GeV}^2$, $9\,{\rm GeV}^2$
and $16\,{\rm GeV}^2$.

\smallskip
\paragraph{3. Double spin asymmetry {\boldmath $A_{TT}$} at PAX.}
The estimates for the double spin asymmetry $A_{TT}$ as defined
in Eq.~(\ref{Eq:ATT-1}) for the PAX kinematics on the basis of
the ingredients discussed above is shown in Fig.~1a. The
exploitable rapidity range shrinks with increasing dilepton mass
$Q^2$. Since $s=x_1x_2Q^2$, for $s=45\,{\rm GeV}^2$ and
$Q^2=5\,{\rm GeV}^2$ ($16\,{\rm GeV}^2$) one probes parton
momenta $x>0.3$ ($x>0.5$). The asymmetry $A_{TT}$ grows with
increasing $Q^2$ where larger parton momenta $x$ are involved,
since $h_1^u(x)$ is larger with respect to $f_1^u(x)$ in the
large $x$-region.

\begin{figure}[h!]
\begin{center}
\vspace{-7mm}
\epsfxsize=5cm\epsfbox{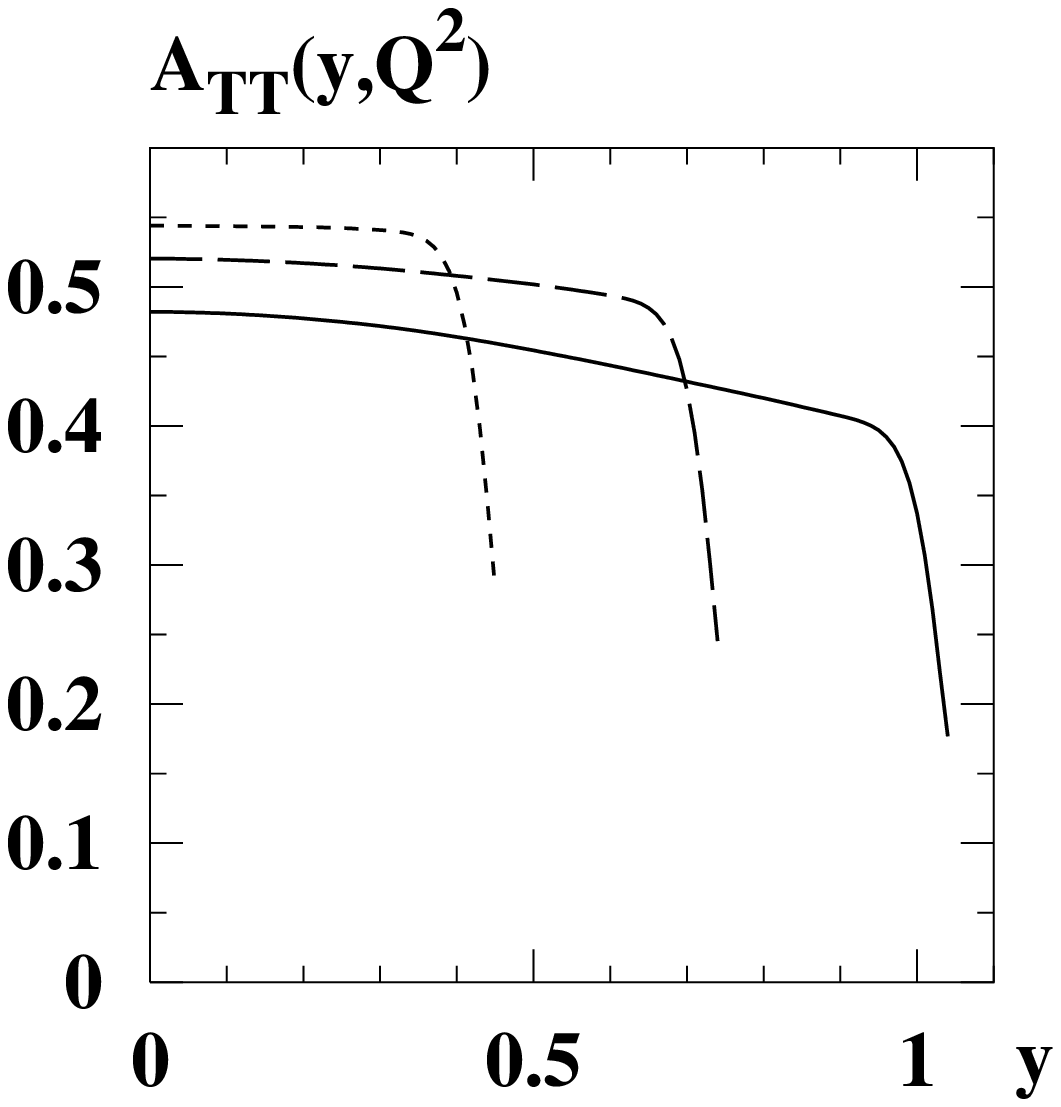} {\bf a)}
\epsfxsize=5cm\epsfbox{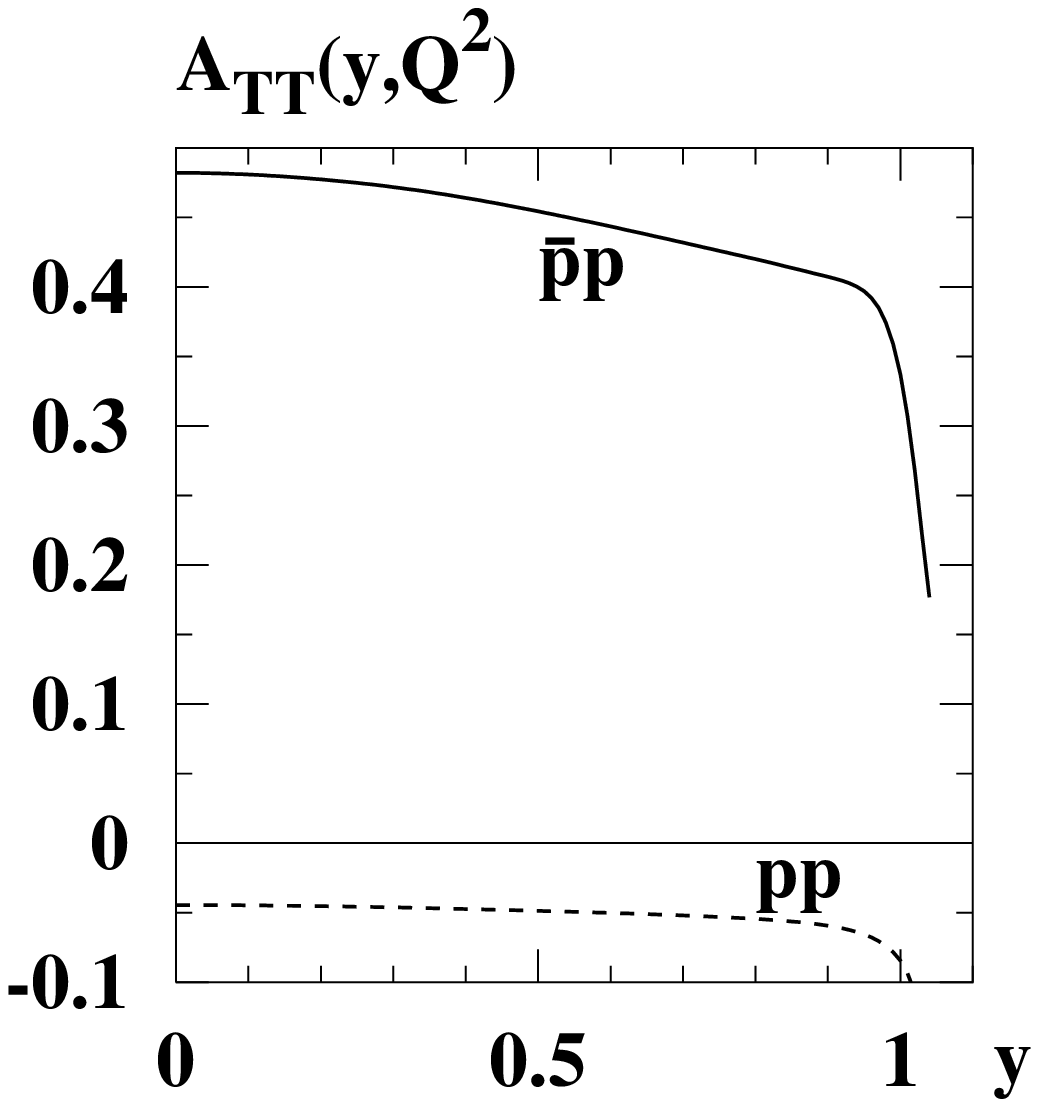} {\bf b)}
\end{center}
\caption{
{\bf a)}
The asymmetry $A_{TT}(y,M^2)$, {\sl cf.} Eq.~(\ref{Eq:ATT-1}),
as function of the rapidity $y$ for $Q^2=5\,{\rm GeV}^2$ (solid)
and $9\,{\rm GeV}^2$ (dashed) and $16\,{\rm GeV}^2$ (dotted line)
for $s=45\,{\rm GeV}^2$.
{\bf b)}
Comparison of $A_{TT}(y,M^2)$ from proton-antiproton
(solid) and proton-proton (dotted line) collisions at PAX for
$Q^2=5\,{\rm GeV}^2$ and $s=45\,{\rm GeV}^2$.
}
\end{figure}

\vspace{-5mm}
The advantage of using antiprotons is evident from Fig.~1b. The
corresponding asymmetry from proton-proton collisions is an order
of magnitude smaller (this observation holds also in the
kinematics of RHIC\cite{Schweitzer:2001sr}). At first glance this
advantage seems to be compensated by the polarization factor in
Eq.~(\ref{Eq:ATT-0}). For the antiproton beam polarization of
$(5-10)\%$ and the proton target polarization of $90\%$, i.e.\ at
PAX $D_P\approx 0.05$. However, thanks to the use of antiprotons
the counting rates are more sizeable. A precise measurement of
$A_{TT}$ in the region $Q>4\,{\rm GeV}$ is very difficult,
however, in the dilepton mass region below the $J/\Psi$
threshold\cite{PAX} and in the resonance
region\cite{Anselmino:2004ki} $A_{TT}$ could be measured with
sufficient accuracy in the PAX experiment.

A precise measurement would allow to discriminate between
different models for $h_1^a(x)$. E.g., on the basis of the
non-relativistic quark model motivated popular guess
$h_1^a(x)\approx g_1^a(x)$ (at some unspecified low scale) one
would expect\cite{Anselmino:2004ki} an $A_{TT}$ of about $30\%$
to be contrasted with the chiral quark soliton model estimate of
about $50\%$.

At next-to-leading order in QCD one can expect corrections to
this result which reduce somehow the asymmetry\cite{NLO}.
Similarly large asymmetries can be also expected in the recently
suggested process of lepton pair production via $\jipsi$
production\cite{Anselmino:2004ki}.

\end{document}